\documentclass[12pt,onecolumn]{IEEEtran}

\usepackage{amssymb}
\usepackage{graphicx}
\usepackage{amsmath}
\usepackage[colorlinks,linkcolor=blue,citecolor=red,linktocpage=true]{hyperref}

\begin {document}


\def\bbbf{{\rm I\!F}}

\def\bbbz{{\mathchoice {\hbox{$\sf\textstyle Z\kern-0.4em Z$}}
{\hbox{$\sf\textstyle Z\kern-0.4em Z$}}
{\hbox{$\sf\scriptstyle Z\kern-0.3em Z$}}
{\hbox{$\sf\scriptscriptstyle Z\kern-0.2em Z$}}}}

\newtheorem{theorem}{Theorem}
\newtheorem{lemma}{Lemma}
\newtheorem{corollary}{Corollary}

\newtheorem{example}{Example}


\title{Determining the full weight distribution of any irreducible cyclic code over any finite field of dimension two}

\author{
Gerardo Vega\thanks{G. Vega is with the Direcci\'on General de C\'omputo y de Tecnolog\'{\i}as de Informaci\'on y Comunicaci\'on, Uni\-ver\-si\-dad Nacional Aut\'onoma de 
M\'exico, 04510 Ciudad de M\'exico, MEXICO (e-mail: gerardov@unam.mx).}\thanks{Manuscript partially supported by PAPIIT-UNAM IN107515.}}
\maketitle


\begin{abstract} 
In coding theory, a very interesting problem (but at the same time, a very difficult one) is to determine the weight distribution of a given code. This problem is even more interesting for cyclic codes, and this is so, mainly because they possess a rich algebraic structure, which can be utilized in a variety of ways. For example, in Engineering Telecommunications, this structure is commonly employed to design very efficient coding and decoding algorithms. In this note, we are going to use a characterization of all semiprimitive two-weight irreducible cyclic codes over any finite field, recently presented in \cite{Vega2}, in order to determine the full weight distribution of any irreducible cyclic code over any finite field of dimension two. In fact, the relevance of our results is that by means of them we can actually directly determine the full weight distribution of any irreducible cyclic code, of dimension one or two, just by knowing its length. 
\end{abstract}


\section{Introduction}
It is said that a cyclic code is irreducible if its parity-check polynomial is irreducible. Over a number of years, several authors have dedicated their efforts to solving the problem of determining the weight distribution of families of irreducible cyclic codes (see for example \cite{Helleseth}, \cite{Klove}, \cite{Schmidt}, \cite{Wolfmann}, \cite{Vega1}, \cite{Ding} and \cite{Sharma}), and this has been so because the weight distribution determines the capabilities of error detection and correction of a given code. On the other hand, the family of cyclic codes is important because it possesses a rich algebraic structure that can be utilized in a variety of ways, particularly, in the design of very efficient coding and decoding algorithms. There are instances where the problem of determining the weight distribution of an irreducible cyclic code is quite simple. For example, it is not difficult to see that any irreducible cyclic code over $\bbbf_{q}$, of length $n$ and dimension one, is equivalent to a repetition code of length $n$, and therefore its weight enumerator polynomial is always of the form: $1+(q-1)z^n$. However, as was pointed out in \cite{Ding} the problem of determining the weight distributions of an irreducible cyclic code is, in general, notoriously difficult. The purpose of this note is to employ the characterization of all one-weight irreducible cyclic codes and the characterization of all semiprimitive two-weight irreducible cyclic codes, presented, respectively, in \cite{Vega1} and \cite{Vega2}, in order to show that any irreducible cyclic code of dimension two is either a one-weight irreducible cyclic code or a semiprimitive two-weight irreducible cyclic code. Since the weight distributions of these two types of irreducible cyclic codes are already known, we can then determine the full weight distribution of any irreducible cyclic code over any finite field of dimension one or two. Furthermore, as will be shown below, the relevance of our results in this note, is that by means of them we can actually directly determine the full weight distribution of any irreducible cyclic code, of dimension one or two, just by knowing its length.

This note is organized as follows: In Section \ref{secdos} we recall the characterization of all one-weight irreducible cyclic codes and the characterization of all semiprimitive two-weight irreducible cyclic codes. In Section \ref{sectres} we use these characterizations in order to determine the full weight distribution of any irreducible cyclic code over any finite field of dimension one or two. Finally, Section \ref{conclusiones} will be devoted to presenting our conclusions.

\section{Two characterizations for two different types of irreducible cyclic codes}\label{secdos}

Throughout this note, we are going to use the following:

\medskip
\noindent
{\bf Notation.} By using $p$, $t$, $q$, $k$ and $\Delta$, we will denote five positive integers such that $p$ is a prime number, $q=p^t$ and $\Delta=(q^k-1)/(q-1)$. Unless otherwise stated, from now on $\gamma$ will denote a fixed primitive element of $\bbbf_{q^k}$, and for any integer $a$, the polynomial $h_a(x) \in \bbbf_{q}[x]$ will denote the {\em minimal polynomial} of $\gamma^{-a}$ (see, for example, \cite[p. 99]{MacWilliams}). Finally, for integers $v$ and $w$, such that $\gcd(v,w)=1$, $\mbox{ord}_{v}(w)$ will denote the {\em multiplicative order} of $w$ modulo $v$.

\medskip

An important type of irreducible cyclic codes are the so-called one-weight irreducible cyclic codes, also known as subfield codes. The following result (see \cite{Vega1}, and alternatively, \cite{Vega2}) is a characterization of this type of irreducible cyclic codes. 

\begin{theorem}\label{teouno}
With our current notation, let $a$ be any integer. Also let $u$ and $n$ be integers in such a way that $u=\gcd(\Delta,a)$ and $n=\frac{q^k-1}{\gcd(q^k-1,a)}$. Assume that $\deg(h_a(x))=k$. Then, $h_a(x)$ is the parity-check polynomial of an $[n,k]$ one-weight irreducible cyclic code over $\bbbf_{q}$, whose nonzero weight is $\frac{n}{\Delta} q^{k-1}$, if and only if $u=1$.
\end{theorem}

Another important type of irreducible cyclic codes are the so-called two-weight irreducible cyclic codes. This type of irreducible cyclic codes were characterized in \cite{Schmidt}, and, as an indirect result, the authors also showed that it is possible to characterize all semiprimitive two-weight irreducible cyclic codes over any finite field. Given the importance of such indirect result, the characterization of all semiprimitive two-weight irreducible cyclic codes was formally presented in \cite[Theorem 7]{Vega2}. As will be clear later, this characterization is of main importance for this note, and therefore we are going to recall such result by means of the following:

\medskip

\begin{center}
TABLE I \\
{\em Weight distribution of a semiprimitive two-weight code ${\cal C}$. \\}
Here $s=(k t)/f$, where $f=\mbox{ord}_{u}(p)$ and $u=\gcd(\Delta,a)$.
\end{center}
\begin{center}
\begin{tabular}{|c|c|} \hline
{\bf Weight} & $\;$ {\bf Frequency} $\;$\\ \hline \hline
0 & 1 \\ \hline
$\; \frac{n q^{k/2-1}}{\Delta}(q^{k/2} - (-1)^s ) \;$ & $\frac{(q^k-1)(u-1)}{u}$ \\ \hline
$\; \frac{n q^{k/2-1}}{\Delta}(q^{k/2} + (-1)^s (u-1) ) \;$ & $\frac{(q^k-1)}{u}$ \\ \hline
\end{tabular}
\end{center}

\medskip

\begin{theorem}\label{teodos}
Consider the same notation and assumption as in Theorem \ref{teouno}. Define $f=\mbox{ord}_{u}(p)$ (observe that $\gcd(p,u)=1$). Then $h_a(x)$ is the parity-check polynomial of a semiprimitive two-weight irreducible cyclic code, ${\cal C}$, if and only if $u=2$ or $u > 2$, $f$ is even and $p^{f/2} \equiv -1 \pmod{u}$. In addition, if ${\cal C}$ is a semiprimitive two-weight irreducible cyclic code, then ${\cal C}$ is an $[n,k]$ cyclic code over $\bbbf_{q}$, with the weight distribution given in Table I. 
\end{theorem} 

We end this section with the following simple result, which basically is just a bridge that will allow us to use Theorem \ref{teodos} in order to determine the full weight distribution of any irreducible cyclic code over any finite field of dimension two.

\begin{lemma}\label{lemauno}
With our current notation, let $u$, $f$ and $s$ be positive integers such that $u \geq 2$, $u | (q+1)$, $f=\mbox{ord}_{u}(p)$, and $s=(2t)/f$. Then $f$ is even and $p^{f/2} \equiv -1 \pmod{u}$, if $u>2$. In addition, we also have that $s$ is even if and only if $u=2$.
\end{lemma}

\begin{proof}
If $u=2$, clearly $f=1$, and hence, $s$ is even. Suppose now that $u>2$. Since $q=p^t \equiv -1 \pmod{u}$, we take $v$ to be the smallest positive integer, such that $p^v \equiv -1 \pmod{u}$. Consequently, we have $f=2v$ and $t=vw$, for some odd integer $w$. Therefore, $f$ is even, $p^{f/2} \equiv -1 \pmod{u}$ and $s=w$. 
\end{proof}

\section{The full weight distribution for any irreducible cyclic code of dimension one or two}\label{sectres}

We now present our main result by means of the following:

\begin{theorem}\label{teotres}
With our current notation, let $\gamma$ be a primitive element of $\bbbf_{q^2}$, and let $a$ be an integer in such a way that $h_a(x) \in \bbbf_{q}[x]$ is the minimal polynomial $\gamma^{-a}$. Fix $u=\gcd(q+1,a)$ and $n=\frac{q^2-1}{\gcd(q^2-1,a)}$. In addition, let ${\cal C}$ be the irreducible cyclic code over $\bbbf_{q}$ of length $n$, whose parity-check polynomial is $h_a(x)$. Then, $u=q+1$ if and only if $\deg(h_a(x))=1$. In addition, the following conditional statements are true:

\begin{enumerate}
\item[(A)] If $u=1$, then ${\cal C}$ is an $[n,2]$ one-weight irreducible cyclic code, whose nonzero weight is $\frac{n}{q+1} q$.

\item[(B)] If $2 \leq u < q+1$, then ${\cal C}$ is an $[n,2,\frac{n(q+1-u)}{q+1}]$ semiprimitive two-weight irreducible cyclic code whose weight enumerator polynomial is: 

\[ 1+\frac{(q^2-1)}{u}z^{\frac{n(q+1-u)}{q+1}}+\frac{(q^2-1)(u-1)}{u}z^{n} \; . \] 

\item[(C)] If $u=q+1$, then ${\cal C}$ is an $[n,1]$ one-weight irreducible cyclic code, whose nonzero weight is $n$ (this code is equivalent to a repetition code of length $n$).

\end{enumerate}
\end{theorem}

\begin{proof}
Clearly, $u=q+1 \Leftrightarrow (q^2-1) | a(q-1) \Leftrightarrow \deg(h_a(x))=1$. Part (A) comes directly from Theorem \ref{teouno}. Now, if $2 \leq u < q+1$, and because $u | (q+1)$, Part (B) comes as direct applications of Lemma \ref{lemauno} and Theorem \ref{teodos}. Part (C) is trivial.
\end{proof}

\begin{example}
Let $q=5$. Thus $q+1=6=2\cdot3$, and because $u=\gcd(q+1,a)$, for an integer $a$, we have that all possible values for $u$ are 1,2,3 and 6. Then, with this easy-to-obtain information, and by means of Theorem \ref{teotres}, we can see that all possible weight enumerator polynomials for an irreducible cyclic code ${\cal C}$ over $\bbbf_{5}$, of dimension one or two, are given by means of the following:
\end{example}

\begin{center}
TABLE II \\
{\em Possible weight enumerator polynomials for ${\cal C}$. \\}
Here $h_a(x)$ is the minimal polynomial of $\gamma^{-a}$ and the parity-check polynomial of the cyclic code ${\cal C}$.
\end{center}
\begin{center}
\begin{tabular}{|c|c|c|c|c|} \hline
$\begin{array}{c} a \; \mbox{ belonging to the} \\ \mbox{cyclotomic coset:} \end{array}$ & $\deg(h_a(x))$ & $\; u \;$ & $\; n \;$ & $\begin{array}{c} \mbox{Weight enumerator} \\ \mbox{polynomial of } {\cal C} \end{array}$ \\ \hline \hline
$\begin{array}{c} \{1,5\}, \{7,11\}, \\ \{13,17\} \;\: \mbox{\em or }\: \{19,23\} \end{array}$ & 2 & 1 & 24 & $1+24z^{20}$ \\ \hline
$\{2,10\} \;\: \mbox{\em or }\: \{14,22\}$ & 2 & 2 & 12 & $1+12z^{8}+12z^{12}$ \\ \hline
$\{4,20\}$ & 2 & 2 & 6 & $1+12z^{4}+12z^{6}$ \\ \hline
$\{8,16\}$ & 2 & 2 & 3 & $1+12z^{2}+12z^{3}$ \\ \hline
$\{3,15\} \;\: \mbox{\em or }\: \{9,21\}$ & 2 & 3 & 8 & $1+8z^{4}+16z^{8}$ \\ \hline
$\{6\} \;\: \mbox{\em or }\: \{18\}$ & 1 & 6 & $4$ & $1+4z^{4}$ \\ \hline
$\{12\}$ & 1 & 6 & $2$ & $1+4z^{2}$ \\ \hline
$\{0\}$ & 1 & 6 & $1$ & $1+4z$ \\ \hline
\end{tabular}
\end{center}

\medskip

Now observe that if the length $n$ in the previous theorem is known, instead of the integer $a$, then $u=\gcd(q+1,\frac{q^2-1}{n})$. Thus, with this in mind, we can now determine the full weight distribution of any irreducible cyclic code, of dimension one or two, just by knowing its length $n$. We formally state this result by means of the following:

\begin{corollary}\label{coruno}
With our current notation, let $n$ be a divisor of $q^2-1$, and $u=\gcd(q+1,\frac{q^2-1}{n})$. If ${\cal C}$ is an irreducible cyclic code over $\bbbf_{q}$, of length $n$, then its full weight distribution is determined by means of the conditional statements in Theorem \ref{teotres}.
\end{corollary}

\begin{example}
Let $q=27$ and $n=104$. Then $u=7$ and, thanks to the previous corollary, the weight enumerator polynomial of an irreducible cyclic code over $\bbbf_{27}$, of length $104$ is: $1+104z^{78}+624z^{104}$.
\end{example}

\section{Conclusion}\label{conclusiones}
In this note, we showed that an irreducible cyclic code of dimension one or two is either a one-weight irreducible cyclic code, of dimension one or two, or a semiprimitive two-weight irreducible cyclic code. Since the weight distributions of this two types of irreducible cyclic codes are already known, we were able to determine the full weight distribution of any irreducible cyclic code over any finite field of dimension one or two. What we consider as an interesting result for this note, is that by Corollary \ref{coruno} it is now possible to directly determine the full weight distribution of any irreducible cyclic code of dimension one or two, just by knowing its length.

\bibliographystyle{IEEE}

\end{document}